\documentclass[preprint,12pt,authoryear]{elsarticle}

\usepackage{amssymb}
 \usepackage{amsthm}

\usepackage{diagbox}
\usepackage{bm}
\usepackage{subfig}
\usepackage{amsmath}

\newtheorem{theorem}{Theorem}[section]

\newtheorem{lemma}[theorem]{Lemma}

\newcommand{\be}{\bm{e}}
\newcommand{\bX}{\bm{X}}

\newcommand{\bv}{\bm{v}}
\newcommand{\balpha}{\bm{\alpha}}
\newcommand{\btheta}{\bm{\theta}}
\newcommand{\bxi}{\bm{\xi}}
\newcommand{\bW}{\bm{W}} 

\graphicspath{{figures/}}

\begin{document}

\begin{frontmatter}



\title{Robust bent line regression}

 \author[label1,label2]{Feipeng Zhang}
 \address[label1]{Department of Statistics, Hunan University, Changsha, 410082, China}

\author[label2]{Qunhua Li\corref{cor1}}
\address[label2]{Department of Statistics, Pennsylvania State University,  PA, 16802, USA}
\cortext[cor1]{Department of Statistics, Pennsylvania State University,  PA, 16802, USA}
\ead{qunhua.li@psu.edu}

\begin{abstract}
We introduce a rank-based bent linear regression with an unknown change point.  
Using a linear reparameterization technique,  we propose a rank-based estimate that can make simultaneous inference on all model parameters, including the location of the change point, in a computationally efficient manner.  We also develop a   score-like test for the existence of a change point, based on a weighted CUSUM process. This test only requires fitting the model under the null hypothesis in absence of a change point, thus it is computationally more efficient than likelihood-ratio type tests. The asymptotic properties of the test are derived under both the null and the local alternative models. Simulation studies and two real data examples show that the proposed methods are robust against outliers and heavy-tailed errors in both parameter estimation and hypothesis testing. 

\end{abstract}


\end{frontmatter}



\section{Introduction}
\label{s:intro}


Segmented linear regression is commonly used for dealing with data in which the relationship between response and explanatory variables is approximately piecewise linear.  
Such data can be encountered in many applications in  medical research, biology, ecology,  insurance and finance studies. For example, in hydrologic studies, the transportation of particles in gravel bed streams is often described as  occurring in phases, with a relatively stable transport rate at low discharge, and a drastic increase after the discharge passes a certain threshold 
\citep{ryan2002defining}.  
Another example arises from a study of the maximal running speed (MRS) data of land mammals  \citep{garland1983relation},   
which shows that the logarithm of MRS increases stably with the logarithm  of the body mass, and gradually decreases  after reaching a certain point.  
The common feature between these examples is that the response and the covariate of interest show a piecewise linear relationship that has varying slopes over different domains of the covariate. 
Besides estimating the regression coefficients, identifying the threshold at which a change of relationship occurs is also a primary interest in statistical analyses. 

In this article, we focus on an important special case of segmented linear regression: the so-called bent line regression.  
This type of regression model comprises of two line segments with different slopes intersecting at a change point, and is used for modelling data with a continuous segmented relation. The two examples mentioned above demonstrate such a relation.
As the location of the change point is unknown, 
 the likelihood function of this model is non-differentiable with respect to the location of the change point, complicating parameter estimation and statistical inference.
Many works have been done to estimate parameters for bent line regression models with normally distributed responses, for example, 
\citet{quandt1958estimation, quandt1960tests}, 
\citet{sprent1961some}, 
\citet{hinkley1969inference}, 
\citet{feder1975asymptotic}, 
\citet{gallant1973fitting}, 
\citet{chappell1989fitting}, 
and many others. 
Most of these methods are based on the grid-search approach \citep{lerman1980fitting}, 
which estimates the regression coefficients for a series of fixed change points on a grid, and then exhaustively searches for the point that maximizes the likelihood function. 
While generating reasonable estimates, 
this approach is computationally expensive and the statistical inference of its estimators is difficult to derive.
Recently,  
\citet{muggeo2003estimating} 
proposed a clever estimation method for this model. By using a simple linearization technique, this method allows simultaneous inference for all model parameters in a computationally efficient manner. 

Although the aforementioned models work well when normality holds, datasets in real applications often have outliers or  heavy-tails,  which can substantially influence the fitting of the models  and the accuracy of parameter estimation.  
For instance, the MRS data includes several extremely slow outliers, which are animals living in environments where high running speed does not give a selective advantage, for example, sloths. The relationship between body mass and running speed for these animals is drastically different from that for most animals living in enviroments where speed is important. 
Even though these animals contribute little information towards the understanding of how body mass affects the maximal running speed, they markedly influence the estimation results.      
In such situations, a robust estimation procedure usually is desirable.  
A common way to obtain robust estimates is  rank-based regression. 
Rank-based regression makes no assumption on the distribution of the response. It is robust against outliers and heavy-tailed errors,  while maintaining high efficiency. 
The inference for rank-based regression models, in absence of change points, has been well developed since the first work by 
\citet{jureckova1971nonparametric} 
and \citet{jaeckel1972estimating},  
see  \citet{abebe2001rank}, \citet{hettmansperger2011robust},  
and the references therein. 
However, to the best of our knowledge,  
no analogous work has been done when a change point is involved.  

In this article, 
we develop a robust rank-based estimate for the bent line regression model with an unknown change-point.  
The main idea of our estimation procedure is to replace the residual sum of squares in the segmented procedure of 
\citet{muggeo2003estimating} 
with the rank dispersion function in standard rank-based regressions 
\citep{jaeckel1972estimating}.  
As a result, it not only achieves robustness against outliers and heavy-tailed errors, but also inherits the merit of Muggeo's segmented method, providing simultaneous estimation and inference  for all model parameters, including the location of the change point.  
It can be implemented readily using  existing packages for standard rank-based regression.  

To ensure the identifiability of the change point in the estimation procedure, we also develop
a testing procedure for the existence of a change point.
Many tests have been developed on determining the existence of a change point in linear regression \citep{andrews1993tests, bai1996testing, hansen1996inference},  
quantile regression 
\citep{qu2008testing, li2011bent, aue2014segmented, zhang2014testing}, 
transformation models 
\citep{kosorok2007inference}, 
time series models 
\citep{chan1993consistency, cho2007testing},  
among others. 
Recently, 
a general method for detecting the structural change in a continuous threshold covariate was proposed by  
\cite{lee2011testing}, by using sup-quasi-likelihood-ratio type statistics. 
 Our test is motivated from the test for structural
change in regression quantiles \citep{qu2008testing}.  
It is a weighted CUSUM type statistic based on sequentially evaluated subgradients for a subsample. It only requires fitting the model under the null hypothesis in absence of a change point,  
thus it is computationally more efficient than likelihood-ratio type tests,  which require fitting both null and alternative hypotheses.    
The limiting distributions of the proposed test statistic under both the null and local alternative models are derived, and the implementation procedures are provided.   

The rest of the article is organized as follows. 
Section~\ref{s:method} introduces the main methodology, including the rank-based estimation procedure and the test for the existence of a change point.   
Sections~\ref{s:simulation} and \ref{s:applications} evaluate the performance of the proposed estimate using simulation studies and two real data examples, respectively.  
Section~\ref{s:discuss} provides the conclusion with possible future enhancement.  

\section{Methodology.}
\label{s:method}
\subsection{Robust bent line regression model}
\label{subs:est}

Let $\{(Y_i, \bX_i, Z_i),~i=1,\cdots, n\}$ be a sample of $n$ independent and identically distributed observations,  
where $Y_i$ is the response variable, 
$\bX_i$ is a $p\times 1$ vector of linear covariates, 
$Z_i$ is a scalar covariate whose relationship with $Y_i$ changes at a change-point location.  
To capture the linear relationship between the response $Y_i$ and the covariates $\bX_i$, and 
the segmented relationship between the response $Y_i$ and the explanatory variable $Z_i$,  
we consider the piecewise linear model
\begin{align}\label{mod1}
	Y_i = \balpha^\top \bX_i  + \beta Z_i + \gamma(Z_i-\tau)_+ + e_i,\quad i=1,...,n, 
\end{align}
where $\btheta=(\balpha^\top, \beta, \gamma)^\top$ are unknown coefficients, $\tau$ is the change point,  $(Z_i-\tau)_+=\max(Z_i-\tau,0)=(Z_i-\tau)I(Z_i>\tau)$,  
and $e_i$ are independent and identical random errors with an unknown distribution $F(\cdot)$. 
The  vector $\balpha$ is the linear regression coefficients for $\bX_i$,  
the scalar $\beta$ is the slope relating $Y_i$ to $Z_i$ for the segment before the change point,  
and $\gamma$ is the difference in slope between the segments before and after the change point.  
It is commonly assumed $\gamma\neq 0$ for identifiability of $\tau$ in model (\ref{mod1}).  
In Section ~\ref{subs:test}, we develop a formal test for this assumption. 

As discussed in Introduction, many existing methods  for model (\ref{mod1}) assume $Ee_i=0$ and $Var(e_i) < \infty$, see \citet{quandt1958estimation}, \citet{chappell1989fitting}, \cite{muggeo2003estimating}, and therein references. 
Similar to the ordinal least squares, these methods can be very sensitive to outliers.  When the error distribution has extremely heavy tails, such as Cauchy distribution,  the assumption of $E(e_i)=0$ is violated and these methods are not appropriate.  This motivates us to seek a robust regression approach based on rank. 

\subsubsection{Rank-based estimator for bent line regression}

To achieve robustness in the bent line regression model, we consider the rank-based estimator based on Jaechel's dispersion function, which was introduced by
 \citet{jureckova1971nonparametric} 
 and 
 \citet{jaeckel1972estimating} 
in the context of classical linear models without change-points. 
The main idea of rank-based estimation is to replace the Euclidean norm in the objective function of the ordinary least square estimator, 
$\|\be\|_2^2=\mathop{\sum}\limits_{i=1}^n e_i e_i$, 
by a pseudo-norm
\begin{equation}
\|\be\|_{\phi}=\mathop{\sum}\limits_{i=1}^n  \phi\left(\frac{R_i}{n+1}\right)e_i,
\end{equation}
where $\be=(e_1, \ldots, e_n)$ are residues, $R_i$ is the rank of the $i$th residual $e_i$ among all residuals, and $\phi(\cdot)$ is a non-decreasing and  square-integrable score function defined on the unit interval $(0,~1)$ satisfying $\int \phi(u)du=0$ and $\int \phi(u)^2du = 1$. The rank-based estimator then is obtained by minimizing $\|\be\|_{\phi}$, which is also called the dispersion function.  
Comparing with the ordinary least squares estimator, the rank-based estimator achieves robustness by downweighting the contribution of large residuals in the sum of residual square through ranks in the score function.  
Here we obtain the rank-based estimator for model (\ref{mod1}) by minimizing the following dispersion function
\begin{align}\label{loss}
D(\btheta,  \tau)=\sum_{i=1}^{n} \phi\left(\frac{R_i}{n+1}\right)e_i,
\end{align}
where  
$R_i$ is the rank of the $i$th residual $e_i=Y_i-\balpha^\top\bX_i-\beta Z_i - \gamma(Z_i-\tau)_+$. 

The score function typically is selected according to the shape of underlying distribution of the error \citep{hettmansperger2011robust}. 
Some commonly-used score functions include the Wilcoxon score function, 
$\phi(t)=\sqrt{12}(t-0.5)$, 
and the sign scores function, 
$\phi(t)=sgn(t-0.5)$. 
For symmetric and moderately heavy-tailed distributions, the Wilcoxon score function has been shown to yield robust and relatively efficient estimators. 
Hence, we use the Wilcoxon score function here.


\subsubsection{Iterative estimating procedure for the rank-based estimator}

One complication in estimating $(\btheta, \tau)$ is that the objective function $D(\btheta, \tau)$ is not differentiable with respect to $\tau$, since the indicator function $I(Z_i>\tau)$ is not differentiable with respect to $\tau$. A possible solution is to follow a grid-search approach commonly-used for piecewise linear models \citep{quandt1958estimation}, which estimates $\btheta$ for a series of fixed $\tau$ on a grid and then exhaustively searches for $\tau$ that maximizes the likelihood function. However, this approach is computationally intensive, 
 and the asymptotic properties of the change point $\tau$ is difficult to derive.

To circumvent this problem, we adopt the linear reparameterization technique proposed by \citet{muggeo2003estimating}. The main idea is to approximate $(Z_i-\tau)_+$ using the first-order Taylor's expansion, such that $\tau$ can be reparameterized as a coefficient term in a continuous linear model and estimated along with other regression coefficients as in the standard regression.  
Comparing with the grid-search method, this method reduces the computational burden and allows the asymptotic properties of all parameters to be derived easily using standard asymptotic theory.

Specifically, we apply the first-order Taylor's expansion around $\tau^{(0)}$,  
provided that $\tau^{(0)}$ is close to $\tau$:  
\[
(Z_i-\tau)_+ \approx (Z_i-\tau^{(0)})_+ +(-1)I(Z_i>\tau^{(0)}) (\tau-\tau^{(0)}).
\]
Then, model (\ref{mod1}) can be approximated by the following model,  
\begin{align}\label{mod2}
Y_i = \balpha^\top \bX_i+ \beta Z_i+ \gamma  (Z_i-\tau^{(0)})_+ 
+ \eta  [-I\{Z_i>\tau^{(0)}\}]+e_i, 
\end{align}
where  $\eta=\gamma (\tau-\tau^{(0)})$.  
For a given $\tau^{(0)}$, by viewing $(Z_i-\tau^{(0)})_+$ and $-I\{Z_i > \tau^{(0)}\}$ as two new covariates, model (\ref{mod2}) takes the form of the standard linear regression.  The rank-based estimate of  regression  coefficients for model (\ref{mod2}) can be obtained using the standard rank-based estimation as  
\[
(\widehat{\btheta}^{(1)},~ \widehat{\eta}^{(1)})
= \arg\min_{\btheta,~\eta}
\sum_{i=1}^{n} \sqrt{12}\left(\frac{R_i^{(0)}}{n+1}-0.5\right)
e_i^{(0)},		
\] 
where $\widehat{\btheta}^{(1)}
=(\widehat{\balpha}^{(1)}, ~ \widehat{\beta}^{(1)}, ~ \widehat{\gamma}^{(1)})$,  
and  $R_i^{(0)}$ is the rank of the $i$th residual $e_i^{(0)}=  Y_i-\balpha^\top \bX_i-\beta Z_i - \gamma(Z_i-\tau^{(0)})_+-
\eta \left(-I\{Z_i>\tau^{(0)}\}\right)$.
The estimate for change-point $\tau$ can be updated by 
\[
\widehat\tau^{(1)}= \widehat\tau^{(0)}+
\frac{\widehat{\eta}^{(1)}}
{\widehat{\gamma}^{(1)}}.
\]

The iterative algorithm is summarized in Algorithm~1. 

\medskip
\fbox{ 
\begin{minipage}{\dimexpr\textwidth-2\fboxsep-2\fboxrule\relax} 
	\parbox{1\textwidth}{%
		\textbf{Algorithm~1}:
		\begin{center}  
			\begin{description}
				\item[(i)] 
				Initialize parameters:  
				$\widehat{\btheta}^{(0)}=(
				\widehat{\balpha}^{(0)},~\widehat{\beta}^{(0)},~ \widehat\gamma^{(0)})$ 
				and  
				$\widehat\tau^{(0)}$, 
				setting  $\widehat\eta^{(0)}$ with a small value, such as $0.01$. 
				
				\item[(ii)] 
				Fix $\widehat\tau^{(s)}$ at each step $s$,  estimate parameters $\widehat{\btheta}^{(s+1)}=(
				\widehat{\balpha}^{(s+1)},~ \widehat{\beta}^{(s+1)},~ \widehat{\gamma}^{(s+1)})$  and $\widehat{\eta}^{(s+1)}$
				by the rank-based regression estimate for the following linear model: 
				\begin{eqnarray}\label{mod3}
				Y_i = \balpha^\top \bX_i+ \beta Z_i+ \gamma  (Z_i-\tau^{(s)})_+ 
				+ \eta  (-I\{Z_i>\tau^{(s)}\}). 
				\end{eqnarray}
				That is, 
				\[
				(\widehat{\btheta}^{(s+1)},~ \widehat{\eta}^{(s+1)})
				= \arg\min_{\btheta,~\eta}
				\sum_{i=1}^{n} \sqrt{12}\left(\frac{R_i^{(s)}}{n+1}-0.5\right)
				e_i^{(s)},		
				\] 
				where $R_i^{(s)}$ is the rank of the $i$th residual $e_i^{(s)}=  Y_i-\balpha^\top\bX_i-\beta Z_i - \gamma(Z_i-\tau^{(s)})_+ -\eta  \left(-I\{Z_i>\tau^{(s)}\}\right)$ among all residuals $e_1^{(s)}, ..., e_n^{(s)}$.

				\item[(iii)] 
				Update the change-point estimate $\widehat\tau^{(s+1)}$ by 
				\begin{equation}\label{upchpt}
				\widehat\tau^{(s+1)}= \widehat\tau^{(s)}+
				\frac{\widehat{\eta}^{(s+1)}}
				{\widehat{\gamma}^{(s+1)}}. 
				\end{equation}
				
				\item[(iv)] Repeat steps (ii)--(iii) until convergence criterion holds, e.g. 		$\|\widehat{\btheta}^{(s+1)}-\widehat{\btheta}^{(s)}\|_\infty < 10^{-5}$.  
				Here, $\|\bv\|_\infty=\mathop{\max}\limits_j|v_j|$ for any $\bv\in\mathbb{R}^q$.				
			\end{description}
			
		\end{center}  
	} 
\end{minipage}
}

\medskip

\textit{Remark~1.}
By viewing $Y_i$ as the response variable and 
$\bX_i, ~ Z_i, ~ (Z_i-\tau^{(s)})_+,~ (-1)I(Z_i>\tau^{(s)}) $ 
as the explanatory variables, fitting the non-linear and non-differentiable model (\ref{mod1}) is equivalent to iteratively  fitting the standard rank-based linear model (\ref{mod3}).  
This fitting procedure can be easily implemented using the standard rank-based regression and computed using existing software tools, such as \textbf{R} package \textit{Rfit}.

By the standard theory of the rank-based linear regression, the estimated coefficient
$(\widehat{\btheta}, \widehat{\eta})$  has an asymptotically normal distribution \citep{hettmansperger2011robust}. 
Using (\ref{upchpt}), the standard error estimate of the change point estimator $\widehat{\tau}$ can be obtained using its Wald statistics. 
Specifically, the standard error of $\widehat{\tau}$ is given by 
\begin{align}\label{eqse}
\hbox{SE}(\widehat{\tau})=\frac{ \left\{\text{Var}(\widehat{\eta})+ \text{Var}(\widehat{\gamma}) (\widehat{\eta}/\widehat{\gamma})^2
+2(\widehat{\eta}/\widehat{\gamma}) 
\hbox{Cov}(\widehat{\eta}, \widehat{\gamma}) \right\}^{1/2}}{ |\widehat{\gamma}| }. 
\end{align}

When the algorithm converges, $\widehat{\eta}$ is expected to be approximately zero. 
Then (\ref{eqse}) is simply $\hbox{SE}(\widehat{\tau})=\hbox{SE}(\widehat{\eta})/|\widehat{\gamma}|$.   
The $100(1-\alpha)\%$ Wald-based confidence interval is given by 
$$
\left[\widehat{\tau}- z_{\alpha/2}\hbox{SE}(\widehat{\tau}), \,  \widehat{\tau}+ z_{\alpha/2} \hbox{SE}(\widehat{\tau})\right],
$$
where $z_{\alpha/2}$ is the $(1-\alpha/2)$th percentile of the standard normal distribution.

\subsection{Test  the existence of a change-point.} 
\label{subs:test}

Note that the convergence of the iterative algorithm depends on the existence of a threshold effect, i.e.  $\gamma\neq 0$. 
If $\gamma = 0$, the change point $\tau$ is not identifiable and its estimation is ill-conditioned. 
Therefore, it is important to test the existence of a threshold effect in the regression model \eqref{mod1}.

Here, we consider the null and alternative hypotheses
\begin{align*}
	H_0: \gamma=0 \quad \text{for any} \quad 
	\tau \in{\Gamma}
	\quad \text{vs.} \quad 
	H_1: \gamma\neq 0 
	\quad\text{for some}\quad 
	\tau\in \Gamma, 
\end{align*}
where $\Gamma$ is the range set of all $\tau$'s. To construct our test statistic, we take a cumulative subgradient approach that is in spirit similar to the test for structural change in quantile regression in \citet{qu2008testing}. 
The key idea of this approach is to construct the test statistic using sequentially evaluated subgradients of the objective function under $H_0$ for a subsample, 
in a fashion similar to the standard CUSUM test \citep{ploberger1992cusum, bai1996testing}.  
One advantage of this approach is that it is a score-like test statistic that can be obtained by only fitting the null model,  
thus it is computationally more efficient than the sup-quasi-likelihood-ratio statistics in 
\cite{lee2011testing}, 
which requires fitting both the null and alternative models.

Specifically, we define 
\[
	R_n(\tau, ~\widehat{\bxi})= \frac{1}{\sqrt{n}}\sum_{i=1}^{n} \sqrt{12}\left(\frac{R(Y_i
		-\widehat{\bxi}^\top\bW_i)}{n+1}-0.5\right) 
	\left(Z_i-\tau\right) I(Z_i\leq \tau), 
\]
where $\widehat{\bxi}
\equiv
(\widehat{\balpha},~ \widehat{\beta})$ is the estimator of the coefficients $\bxi=(\balpha,~ \beta)$ under  the null hypothesis $H_0$,  
\begin{align}\label{nullest}
\widehat{\bxi}
=\arg\min_{\bxi}
\sum_{i=1}^{n} \sqrt{12}\left(\frac{R(Y_i-\bxi^\top\bW_i)}{n+1}-0.5\right)\left(Y_i-\bxi^\top\bW_i\right),
\end{align}
where $\bW_i= (\bX_i^\top, Z_i)^\top$ are covariates   
and $R(Y_i-\bxi^\top\bW_i)$ is the rank of the $i$th residual $Y_i-\bxi^\top\bW_i$ among all the residuals $(Y_1-\bxi^\top\bW_1,\, \cdots \,, Y_n-\bxi^\top\bW_n)$.   $R_n(\tau, ~\widehat{\bxi})$ is a variant of the negative subgradient of the rank-based objective function \eqref{loss} with respect to $\gamma$ under $H_0$, for the subsample with $Z_i$ up to the threshold $\tau$.  
Intuitively, when there is no bent, $\hat{\xi}$ would be a good estimator for its population value, then the estimated residuals $\hat{e}_i=Y_i - \hat{\xi}^T\mathbf{W}_i$ would be close to 0. Meanwhile,  when  there exists a change point,  $\widehat{\bxi}$  would be significantly different from the true value for some subsamples. 
Consequently,  $\widehat{e}_i$ would have a large absolute value away from zero,  resulting in a large absoulte value of $R_n(\tau, ~\widehat{\bxi})$.  
Since the change point is unknown, we need search all the possible locations. 
Therefore, we propose the test statistic  
\[
	T_n =\sup_{\tau\in \Gamma} \left| R_n (\tau, ~\widehat{\bxi}) \right|. 
\]
This statistic can be viewed as a weighted CUSUM statistic based on the ranks of estimated residuals under the null hypothesis. It is intuitively plausible to reject $H_0$ when $T_n$ is too large.  
This intuition will be formally verified by Theorem \ref{thm1}.  
It implies that $R_n(\tau, \widehat{\bxi})$ converges to a Gaussian process with mean zero, and the size of such a process can be used to test for the existence of a change point.

To derive the large-sample inference for $T_n$,  we consider the local alternative model, 
\begin{align}\label{mod4}
Y_i = \balpha^\top \bX_i  + \beta Z_i + n^{-1/2}\gamma(Z_i-\tau)_+ + e_i,\quad i=1,...,n, 
\end{align}
where $\tau$ is the change-point location and $\gamma\neq 0$. For ease of presentation, we define some notations.   
Denote $F(\cdot)$ and $f(\cdot)$ as the cumulative  distribution function and density function of random error $e$, respectively,  and the scale parameter 
$
	c_\phi= \left\{\int \phi'(F(u)) f(u) dF(u)\right\}^{-1}, 
$
which is presented in  \cite{hettmansperger2011robust}. 
Define $S_{wn} =
n^{-1}\sum_{i=1}^{n}  W_i W_i^T $  and   
$S_w=\mbox{E} \left[W_i W_i^T\right]$,  
\begin{align*}
  S_{1n}(\tau) 
&=
  n^{-1}\sum_{i=1}^n \sqrt{12} f(e_i)W_i (Z_i-\tau) I(Z_i\leq \tau),  \\
  S_1(\tau) 
&=
  \mbox{E}  
   \left[
   \sqrt{12} f(e_i)W_i (Z_i-\tau)I(Z_i\leq \tau)
   \right], \\
  S_{2n} (\tau) 
&=
   n^{-1}\sum_{i=1}^{n} \sqrt{12}\gamma f(e_i)W_i (Z_i-\tau)_+, \\
  S_2(\tau) 
&=
  \mbox{E}  
  \left[
   \sqrt{12}\gamma f(e_i)W_i(Z_i-\tau)_+
  \right], 
\end{align*}
and  
$q(\tau) = c_\phi S_{1}(\tau)^\top S_w^{-1} S_{2}(\tau)$.  

The following theorem is essential to the large-sample  inference for using $T_n$. 
\begin{theorem}\label{thm1}
	Under regular conditions in the Appendix A, for the local alternative model \eqref{mod4},  
	$R_n(\tau)$ has the asymptotic representation 
	\begin{align}\label{rnt}
	R_n(\tau, \widehat{\bxi}) &= 
	\frac{1}{\sqrt{n}}\sum_{i=1}^n \sqrt{12} \left[F(e_i)-0.5\right]
	\left[
	(Z_i-\tau)I(Z_i\leq \tau)- 
	c_\phi S_1(\tau)^\top S_w^{-1}W_i
	\right] \\ \nonumber
	&+ q(\tau) +o_P(1). 
	\end{align} 
	Furthermore, 
	$T_n$ converges weakly to the process  $\mathop{\sup}\limits_\tau\left|R(\tau)+q(\tau)\right|$,
	where $R(\tau)$ is the Gaussian process with mean zero and covariance function 
	\begin{align*}
	W(\tau_1, \tau_2)
	=& 
	\mbox{E}
	\bigg[
	\left\{
	(Z-\tau_1)I(Z\leq \tau_1) -c_\phi S_1(\tau_1)^\top S_w^{-1}W
	\right\} \\
	& \times
	\left\{
	(Z-\tau_2)I(Z\leq \tau_2) -c_\phi S_1(\tau_2)^\top S_w^{-1}W
	\right\}
	\bigg]. 
	\end{align*}
\end{theorem}

\textit{Remark~2.}  
Under the null hypothesis $H_0$,  $q(\tau)$ equals to $0$ for all $\tau$, whereas $q(\tau)$ is a nonzero function of $\tau$ under the local alternative model.  
Thus, the proposed test statistic can distinguish the alternative hypothesis from the null hypothesis. 
This supports the intuitive interpretation of the proposed test statistics for the existence of the change point. 

The following theorem implies that the power of the test statistic $T_n$ approaches 1 under the local alternative model whose order of $\gamma$ is arbitrarily close to $n^{-1/2}$. 

\begin{theorem}\label{thm2}
	Under regular conditions in the Appendix A, 
	for the local alternative model, 
	$$
	Y_i = \balpha^\top \bX_i  + \beta Z_i +  n^{-1/2}a_n\gamma(Z_i-\tau)_+ + e_i,\quad i=1,\cdots,n,
	$$
	for any increasing sequence $a_n \rightarrow \infty$, 
	we have  
	$\mathop{\lim}\limits_{n\rightarrow \infty} P\left(|T_n|\geq t \right)=1$ for any $t>0$. 
\end{theorem}

However, the limiting null distribution of $T_n$ is nonstandard, because the covariance of test statistic $T_n$ involves the estimation for the cumulative distribution function $F(\cdot)$ and the density function $f(\cdot)$ of errors.  
To obtain critical values,  
we use a wild bootstrap method similar to that in  
\citet{he2003lack} 
for quantile regression,   
based on the asymptotic representation of $R_n(\tau)$ in \eqref{rnt}.   
The algorithm is summarized in Algorithm 2. 

\textit{Remark~3.} 
Note that the statistic $R_n^*(\tau)$   (defined in Algorithm~2)  depends on the bandwidth $h$ through the kernel estimator $\widehat{S}_{1n}(\tau)$. 
To choose the optimal bandwidth, 
one can use Silverman's rule of thumb 
\citep{silverman1986density}, 
$h =1.06\widehat{\sigma}n^{-1/5}$, where 
$\widehat{\sigma}$ is the standard deviation of the estimated residual $\widehat{e}_i \, (i=1,...,n)$ under the null hypothesis.    
We also perform a sensitivity analysis to evaluate how the choice of $h$ affects the performance of the proposed test procedure (Section~\ref{subs:simtest}).  

In the Appendix, we prove the following result,  
which implies the validity of the bootstrap resampling scheme.   
\begin{theorem}\label{thm3}
	Under both the null and the local alternative hypotheses, $R_n^*(\tau)$ converges to the Gaussian process $R(t)$ as $n\rightarrow \infty$. 
\end{theorem}

\fbox{
\begin{minipage}{\dimexpr\textwidth-2\fboxsep-2\fboxrule\relax} 
	\parbox{1\textwidth}{%
		\textbf{Algorithm~2}:
		\begin{center} 
			\begin{description}
				\item[1]  
				Generate iid $\{u_1,\cdots, u_n\}$  with $u_i=v_iw_i$, where $v_i$ is generated from   $N(0,1)$, 
				and $w_i$ (independent of $v_i$'s) from $P(w_i=1)=P(w_i=-1)=0.5$.   
				
				\item[2]  
				Calculate the test statistic 
				\begin{align*}
				R_n^*(\tau)
				&= \frac{1}{\sqrt{n}}\sum_{i=1}^{n} u_i\sqrt{12}
				\left[\widehat{F}_n(\widehat{e}_i)-0.5\right]	\\
				& \times \left[
				(Z_i-\tau)I(Z_i\leq \tau)- 
				\widehat{c}_\phi \widehat{S}_{1n}(\tau)S_{wn}^{-1}W_i
				\right],
				\end{align*}
				where 
				$$
				\widehat{S}_{1n}(\tau)= \frac{1}{n}\mathop{\sum}\limits_{i=1}^{n}\sqrt{12}
				\widehat{f}(\widehat{e}_i) W_i(Z_i-\tau)I(Z_i\leq \tau),
				$$ 
				$\widehat{F}_n(\cdot)$ is the empirical distribution function of the  estimated residuals $\widehat{e}_i=Y_i -\widehat{\bxi}^\top \bW_i$ under the null hypothesis, 
				$\widehat{f}(\widehat{e}_i)=n^{-1}\sum\limits_{j=1}^{n}K_h(\widehat{e}_i-\widehat{e}_j)$ is a kernel density estimate for the density function $f(\widehat{e}_i)$,   
				$K_h(\cdot)= K(\cdot/h)/h$,   $K(\cdot)$ is a kernel function, 
				and $h>0$ is a bandwidth.
				Here, $\widehat{c}_\phi$ is the consistent estimator for the scale parameter $c_\phi$, 
				which can be readily obtained from the \textbf{R} package \textit{Rfit}. 
				
				\item[3] 
				Repeat Steps~1--2 with $\hbox{NB}$ times to obtain $T_{n}^{*(1)},\cdots, T_n^{*(\hbox{NB})}$. 
				Calculate the p-value as  
				$\widehat{p}_n =\hbox{NB}^{-1}\mathop{\sum}\limits_{j=1}^{\hbox{NB}} I\{T_n^{*(j)}\geq T_n\}$. 	
			\end{description}
		\end{center}
	}%
\end{minipage}
}%

\section{Simulation studies.} 
\label{s:simulation}

\subsection{Estimation}
\label{subs:simest}
To evaluate the finite sample performance of the proposed estimation procedure 
(Section \ref{subs:est}), we conduct several simulation studies using data generated from the following model:
\[
  Y_i =\beta_0
 + \beta Z_i +\gamma(Z_i-0.5)_+ + e_i, \, i=1,...,n, 
\]
with $Z_i\sim \hbox{Uniform}(-2,~2)$, $\tau=0.5$,  
and $(\beta_0,~ \beta,~ \gamma)=(3,~ 2.5, ~-4)$.  
Three different error distributions are considered: 
(1)  a standard normal distribution; 
(2)  a t-distribution with three degrees of freedom, $t_3$; 
and 
(3) a contaminated standard normal distribution,  with  $10\%$ observations  from a standard Cauchy distribution.  
For each setting, we generate a sample of  $n=200$ independent observations $(Z_i, ~Y_i)$ with 1000 repetitions.

To evaluate the performance of our estimator,  
we assess the accuracy of estimation and the appropriateness of Wald-based confidence intervals, and compare its performance with Muggeo's  method, which was implemented in \textbf{R} package \textit{segmented}. 
The results are summarized  as below (Table~\ref{tab:sim}). 
\begin{enumerate}
	\item 
	When the error term follows a standard normal distribution, 
	both estimators work well and have comparable performance: 
	both estimators are unbiased, the estimated standard errors (ESE) are close to the standard deviations (SD), and the empirical coverage probabilities (CP)  approach the nominal level. 
	The mean square errors (MSE) and the average lengths (AL) of Muggeo's estimators are slightly smaller than those of the proposed estimators. 
	This is not surprising, since rank-based estimators for traditional linear regressions with a normal error can achieve 95\% relative efficiency of the ordinary least squares  \citep{hettmansperger2011robust}. 
	
	\item
	When the error term follows a $t_3$ distribution,  both methods work reasonably well, but our estimators have smaller SDs and MSEs than the Muggeo's estimators.  In addition,  the confidence intervals (CIs) of our estimators are shorter than those of Muggeo's estimators, and the empirical coverage probabilities of our CIs are closer to the nominal level than those of Muggeo's CIs for most estimators. 
		
	\item
	When the error term follows a contaminated standard normal distribution with  $10\%$ contamination from a standard Cauchy distribution, 
	Muggeo's method generates biased estimators with drastically inflated SDs and MSEs. 
	However, our method still provides unbiased estimates, reasonable SDs and MSEs. 
	While Muggeo's CI are unreasonably wide with low empirical coverage probabilities, the empirical coverage probabilities of our CIs are still close to the nominal level, and the lengths of CIs are as reasonable as cases 1 and 2. 
\end{enumerate}

In short, comparing with Muggeo's estimators, our estimators achieve robustness against outliers and  heavy-tailed errors. 

\begin{table}
   \caption{Performance comparison between the proposed estimator and Muggeo's estimator, based on 1,000 simulated samples of 200 observations, for the three error distributions in the simulation studies. }\label{tab:sim}
	\begin{minipage}{\textwidth}
	\begin{scriptsize}
		\begin{center}
			\begin{tabular}{clrrrrlrrrr}
				\hline\hline
				\multicolumn{2}{c}{}&
				\multicolumn{4}{c}{Muggeo} &&  \multicolumn{4}{c}{Proposed}\\
				\cline{3-6}\noalign{}  \cline{8-11}\noalign{}
				Case && $\beta_0$ & $\beta_1$ & $\beta_2$ & $\tau$ && $\beta_0$ & $\beta_1$ & $\beta_2$ & $\tau$ \\
				1 & Bias & 0.013 & 0.012 & -0.024 &    -0.006&& 0.026 & 0.023 & -0.011 & -0.017\\ 
				& SD & 0.136 & 0.128 & 0.311 & 0.086 && 0.152 & 0.135 & 0.316 & 0.090 \\ 
				& ESE & 0.131 & 0.126 & 0.303 & 0.074 && 0.150 & 0.131 & 0.308 & 0.077 \\ 
				& MSE & 0.019 & 0.016 & 0.097 & 0.007 && 0.024 & 0.019 & 0.100 & 0.008 \\ 
				& CP & 0.948 & 0.951 & 0.944 & 0.916 && 0.958 & 0.942 & 0.934 & 0.916 \\ 
				& AL & 0.514 & 0.495 & 1.186 & 0.292 && 0.586 & 0.514 & 1.208 & 0.301 \\ 
				2 & Bias & 0.021 & 0.024 & -0.161 & 0.012 
				&& 0.031 & 0.031 & -0.052 & -0.011 \\ 
				& SD & 0.221 & 0.217 & 0.552 & 0.143 && 0.170 & 0.153 & 0.372 & 0.101 \\ 
				& ESE & 0.218 & 0.209 & 0.514 & 0.121 && 0.176 & 0.161 & 0.378 & 0.093 \\ 
			& MSE & 0.049 & 0.048 & 0.330 & 0.021 && 0.030 & 0.024 & 0.141 & 0.010 \\ 
				& CP & 0.935 & 0.935 & 0.946 & 0.901 && 0.967 & 0.962 & 0.950 & 0.931 \\ 
				& AL & 0.853 & 0.819 & 2.015 & 0.476 && 0.689 & 0.632 & 1.481 & 0.366 \\ 
				3 & Bias & 38.694 & 20.100 & 90.527 & -0.041 
				&& 0.020 & 0.017 & -0.011 & -0.011 \\ 
				& SD & 2383.566 & 1216.455 & 3691.305 & 0.439 
				&& 0.159 & 0.140 & 0.324 & 0.097 \\ 
				& ESE & 5.615 & 3.564 & 7.396 & 0.216 && 0.154 & 0.137 & 0.323 & 0.080 \\ 
				& MSE & $5.678\times 10^7$ & $1.479\times 10^7$ & $1.362\times 10^7$ & 0.195 
				&& 0.026 & 0.020 & 0.105 & 0.009 \\ 
				& CP & 0.928 & 0.931 & 0.938 & 0.820 && 0.944 & 0.934 & 0.940 & 0.903 \\ 
				& AL & 22.009 & 13.970 & 28.993 & 0.848 
				&& 0.604 & 0.538 & 1.265 & 0.314 \\ 
				\hline\\ 
			\end{tabular}
		\end{center}
		
		{\vspace*{-5mm} 
		\scriptsize{
				Muggeo: the Muggeo's segmented estimator;
				Proposed: the proposed estimator;
				Bias: the empirical bias;
				SD: the empirical standard error;
				ESE: the average estimated standard error;
				MSE: the average of estimated mean square error.
				CP: $95\%$ coverage probability;
				AL: the average length of $95\%$ confidence intervals. 
			}}\\
	\end{scriptsize}
	\end{minipage}
\end{table} 

\begin{table}
    \caption{Comparison of the testing procedure based on our estimator and the testing procedure based on Muggeo's estimator for the three error distributions in the simulation studies. Type I error and power are calculated at the significance level of 5\% from 1,000 simulated samples of 200 observations. }
    \label{tab:power}
		\begin{center}
			\begin{tabular}{cc|ccccc}
				\hline\hline
				Case & 
				\diagbox[width=8em,trim=l]{Method}{$\gamma$} & 0     
				& -2    & -1    & 1     & 2     \\
				\hline
				1     & Muggeo   & 0.048 & 1.000     & 0.944 & 0.941 & 1.000     \\
				& Proposed  & 0.048 & 1.000     & 0.924 & 0.913 & 1.000     \\
				2     & Muggeo   & 0.298 & 0.999 & 0.872 & 0.861 & 1.000     \\
				& Proposed  & 0.037 & 0.996 & 0.722 & 0.738 & 0.997 \\
				3     & Muggeo   & 0.497 & 0.997 & 0.878 & 0.884 & 0.996 \\
				& Proposed & 0.027 & 0.836 & 0.626 & 0.602 & 0.831		\\
				\hline
			\end{tabular}
		\end{center}
\end{table}

\subsection{Type I error and power analysis}
\label{subs:simtest} 
We evaluate the type I error and power of the testing procedure in Section \ref{subs:test}. As \citep{muggeo2003estimating} did not provide a test for the existence of a change point, we derive a weighted-CUSUM test statistic for Muggeo's model (see Appendix B). We then compare its performance with our test statistic for the ranked-based bent line regression.
We simulate the data from the same simulation settings as the ones in the previous section,  
with threshold effects at $\gamma=-2,-1,0,1,2$. 
In the testing procedure,  
we use the Epanechnikov kernel $K(u)=3/4(1-u^2) I(|u|\leq 1)$, and set the number of bootstrap $\hbox{NB}=1,000$,  
the bandwidth $h =1.06\widehat{\sigma}n^{-1/5}$, and the nominal significance level at $5\%$.  

As shown in Table~\ref{tab:power}, when the error term follows a standard normal distribution, both tests have type I errors close to the nominal level and have reasonable power.  
However, when the error term is distributed as a $t_3$ distribution or is contaminated with a Cauchy distribution,  
the test based on Muggeo's method is anti-conservative,  
with high power but also drastically inflated type I errors. 
This is mainly because Muggeo's method is based on the ordinary least squares, thus it is sensitive to outliers.  
In contrast, our method maintains the nominal level of Type I errors for all error distributions, while having reasonable power.  

\begin{figure}
	 \center
	\includegraphics[height=2.5in, width=1\textwidth]{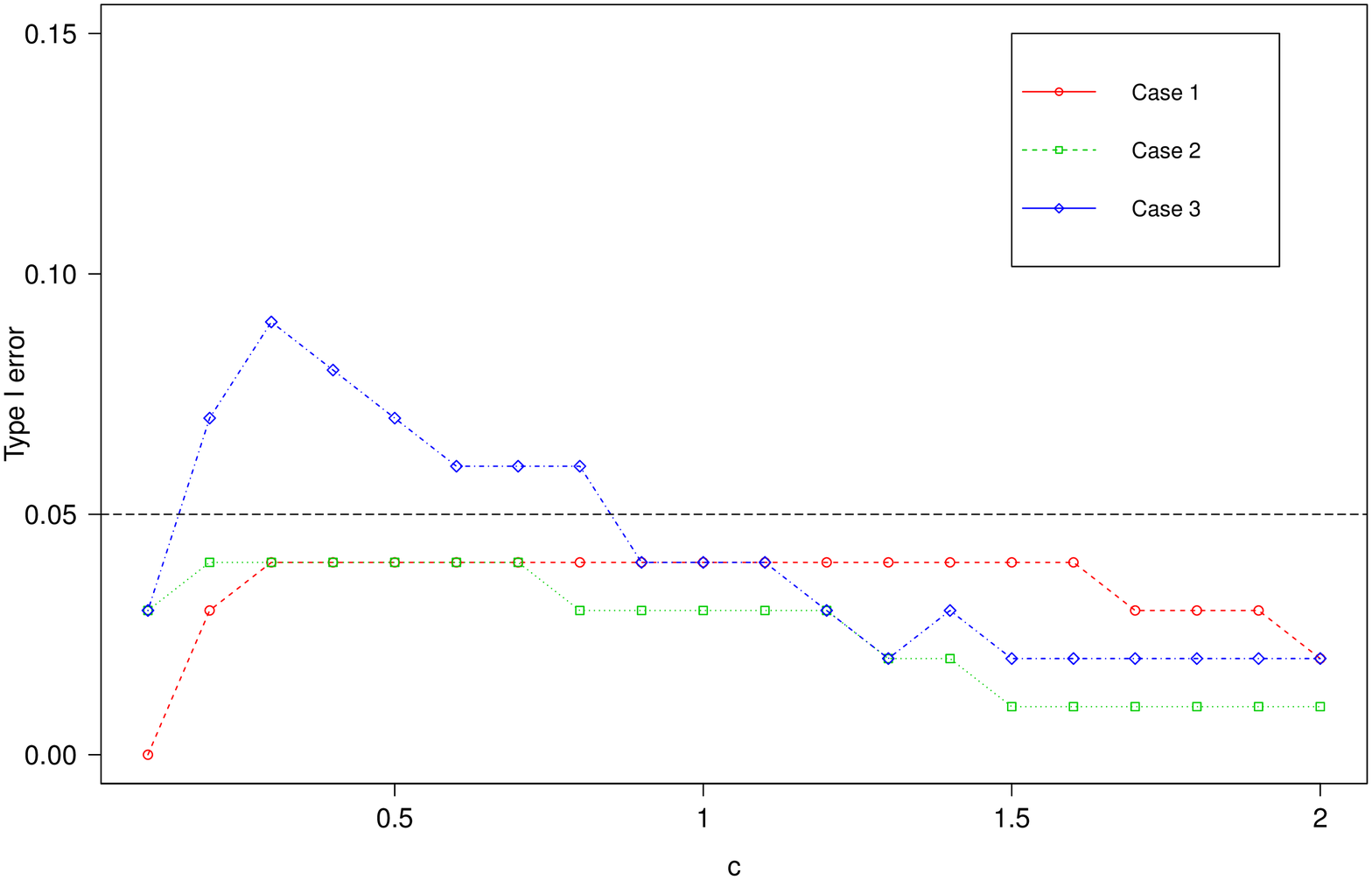}
	\caption{Type I errors of the proposed testing procedure at different bandwidths $h=c \widehat{\sigma} n^{-1/5}$ for the three error distributions in the simulation studies, with $c\in [0.1, 2]$. Each type I error is calculated based on 100 samples of 200 observations at the significant level of 5\%.}
	\label{fig:size}
\end{figure}

We also assess the sensitivity of the proposed method to the choice of bandwidth.  
Here we set the bandwidth as $h=c \widehat{\sigma} n^{-1/5}$, 
and calculate the type I errors at a series of  $c\in [0.1, 2]$ for each error distribution.  
As shown in Figure~\ref{fig:size}, 
the proposed test is not sensitive to the choice of $h$,  
giving reasonable type I errors across a wide range of $c$.  

\section{Applications.} 
\label{s:applications}

\subsection{Bedload transport data}
\label{subs:bed}

In this section, we analyze a bedload transport dataset collected during snow-melt runoff in 1998 and 1999 at Hayden Creek near Salida, Colorado 
\citep{ryan2007tutorial}.  
Bedload transport measures the transportation of  particles in a flowing fluid along the bed.   
In gravel bed streams, bedload transport is generally described as occurring in phases, involving a transition from primarily low rates of sand transport (Phase I) to higher rates of sand and coarse gravel transport (Phase II)
 \citep{ryan2007tutorial}.  
It has been reported that the relationship between transport and water discharge is substantially different in the two phases. The transition of the relationship has been used to define the shift in the phase of transport 
\citep{ryan2002defining}.

In this dataset, the discharge rate ($m^3/s$) and the rate of bedload transport ($kg/s$) were collected for $76$ observations. The dataset has been previously analyzed by 
\citet{ryan2007tutorial}, 
using a piecewise linear regression model. However, as they pointed out, the dataset has very few  observations at higher flows, making it difficult to fit the piecewise linear regression model.  
The {\it loess} curve indeed shows a segmented pattern with a visual estimate of a change point at around $Z_i=1.5m^3/s$.  The two points with the highest transport ($Y_i=0.0536, 0.0673$) are indicated as outliers (p-value$=2.2\times 10^{-16}$) by Grubbs test (Grubbs et al, 1950). 

Here we analyze the dataset using the bent line regression,
\[
Y_i =\alpha
+ \beta Z_i +\gamma(Z_i-\tau)_+ + e_i, \, i=1,...,n, 
\]
where $Z_i$ is the discharge, 
$Y_i$ is the bedload transport rate, 
$\tau$ is the location of the change-point, 
and $e_i$ is the error with unknown distribution. 
Here a change point indicates the discharge at which a phase transition of transport occurs.  

\begin{figure}
	\centering
	\includegraphics[height=2.5in, width = 1\textwidth]{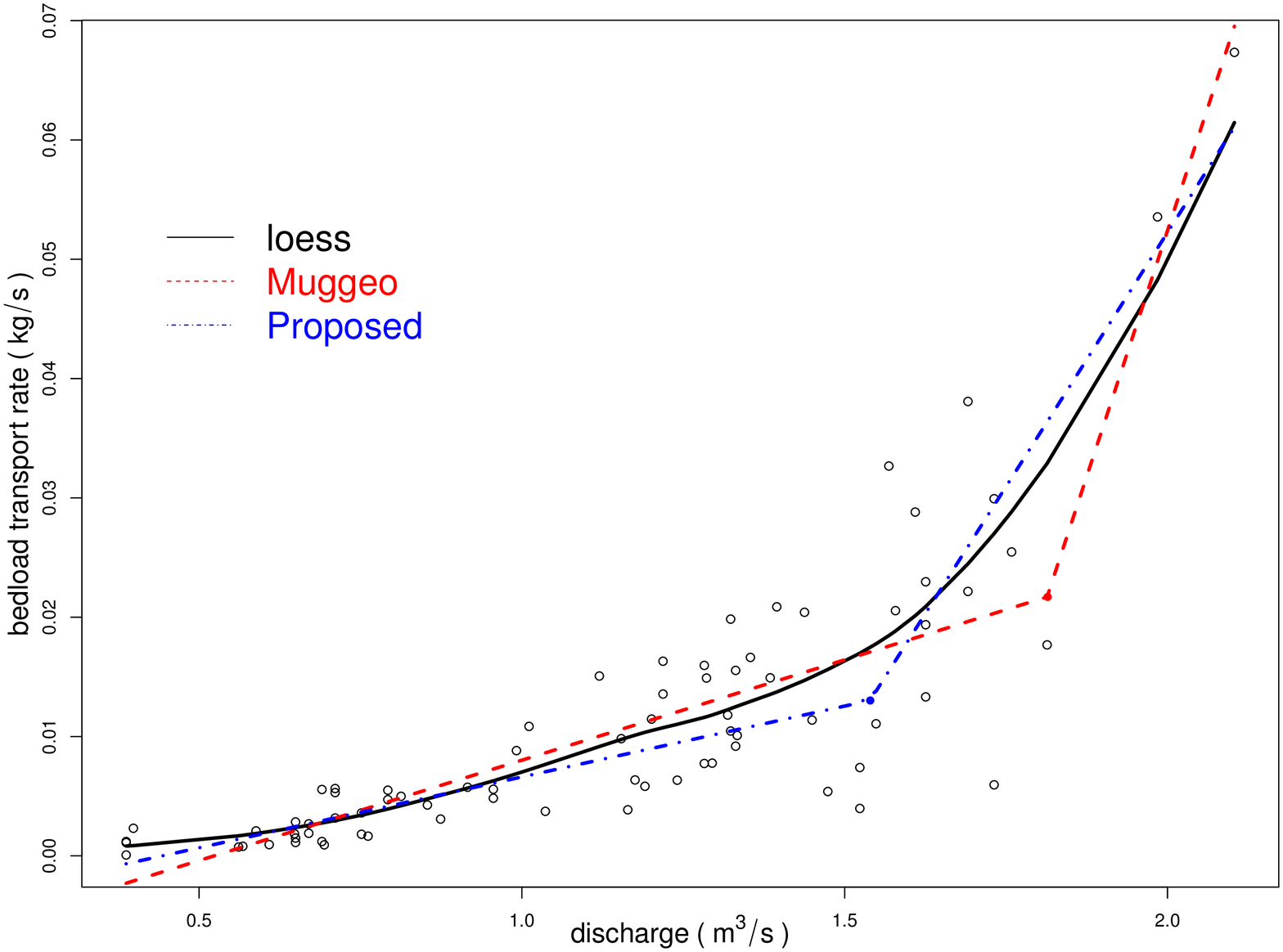}\\
	\caption{Fitted curves for Hayden Creek data, 
		where $``\bullet"$ indicates the location of estimated change-point.}
		\label{fig:hay}
\end{figure}

We first test the existence of a change point using the procedure in Section \ref{subs:test}.  
Our test indicates that the pattern of segmentation is statistically significant (p-value = 0.028). 
Therefore, it is valid to estimate the parameters from the bent line regression model. 
For comparison, we fit the data using Muggeo's method 
\citep{muggeo2003estimating} 
and our method. 
The fitted curves are displayed in Figure~\ref{fig:hay} 
and the estimated parameters are summarized in Table~\ref{tab:hay}. 
For both methods, the fitted line below the change point has a flatter slope with less variability, 
while the line above the change point has a significantly steeper slope and more  variability. 
This reflects the physical characteristics of phases I and II, respectively, 
and is in accordance with the analysis in  \cite{ryan2007tutorial}. 
The estimated change point is $1.813$ by Muggeo's method and $1.539$ by our method. 
Visual inspection of the fitted lines indicates that Muggeo's change point is heavily influenced  by the two outliers, 
whereas our estimate is more robust and is closer to the visual estimate from the \textit{loess} curve. 

To evaluate the performance of model fitting, we use a K-fold cross-validation. 
Specifically, we divide the data into K equal-sized subgroups, 
denoted as $D_k$ for $k = 1, \cdots, K$. 
The $k$th prediction error is given by 
\begin{equation*}
\hbox{PE}_k =\sum_{i\in D_k} \left[Y_i -
\widehat{Y}_i^{(-k)}
\right]^2, 
\end{equation*}
where $\widehat{Y}_i^{(-k)}= \widehat{\alpha}^{(-k)}
+ \widehat{\beta}^{(-k)} Z_i +\widehat{\gamma}^{(-k)}(Z_i-\widehat{\tau}^{(-k)})_+$, 
and parameters $\widehat{\alpha}^{(-k)}$,  
$\widehat{\beta}^{(-k)}$,  
$\widehat{\gamma}^{(-k)}$,  
$\widehat{\tau}^{(-k)}$ are estimated by using the data from all the subgroups other than $D_k$. 
The total prediction error is  $\hbox{PE}=\sum_{k=1}^K\hbox{PE}_k$. 
Here, we set $K = 4$.   
The total prediction error of our method ($0.0038$) is $15.6\%$ less than that of Muggeo's method ($0.0045$).  

\begin{table}
\caption{The estimated parameters  and total prediction errors (PE) for Hayden Creek data.  Their standard errors are listed in parentheses. }\label{tab:hay}
		\begin{center}
			\begin{tabular}{cccccc}
				\hline
				\hline
				& $\alpha$ & $\beta$ & $\gamma$ & $\tau$ & PE\\
				Muggeo  & -0.0088& 0.0168  & 0.1473 & 1.8126 & 0.0045\\
				&(0.0018) & (0.0016) & (0.0636)& (0.1022)\\
				Proposed & -0.0053  & 0.0119  & 0.0733 & 1.5394 & 0.0038\\ 
				&(0.0017)&(0.0016) & (0.0077) & (0.0275)\\
				\hline       
			\end{tabular}
		\end{center}
\end{table}

\subsection{Maximal running speed data} 
\label{subs:mrs}
In this section, 
we analyze the dependency of the maximal running speed (MRS) on body size for land mammals, using a dataset of 107 land mammals collected by \citet{garland1983relation}. 
It is known that the fastest mammals are neither the largest nor the smallest,  
so the dependency is non-monotonic.  
To model this dependency, 
\citet{huxley1936terminology} 
introduced an  allometric equation, 
\begin{equation*}
MRS = \exp(a)\times mass^b,
\end{equation*}
where constants $a$ and $b$ may vary after the mass exceeds some change point. 
This suggests a linear relationship between log(MRS) and log(mass) with a possible change point \citep{chappell1989fitting, li2011bent}.  

\begin{figure}
\centering
\subfloat[][Fitted curves for MRS data, where $``\bullet"$ indicates the location of the estimated change-point, ``h" indicates hoppers and "o" indicates non-hoppers.]{
\includegraphics[height=3in, width = 0.9\textwidth]{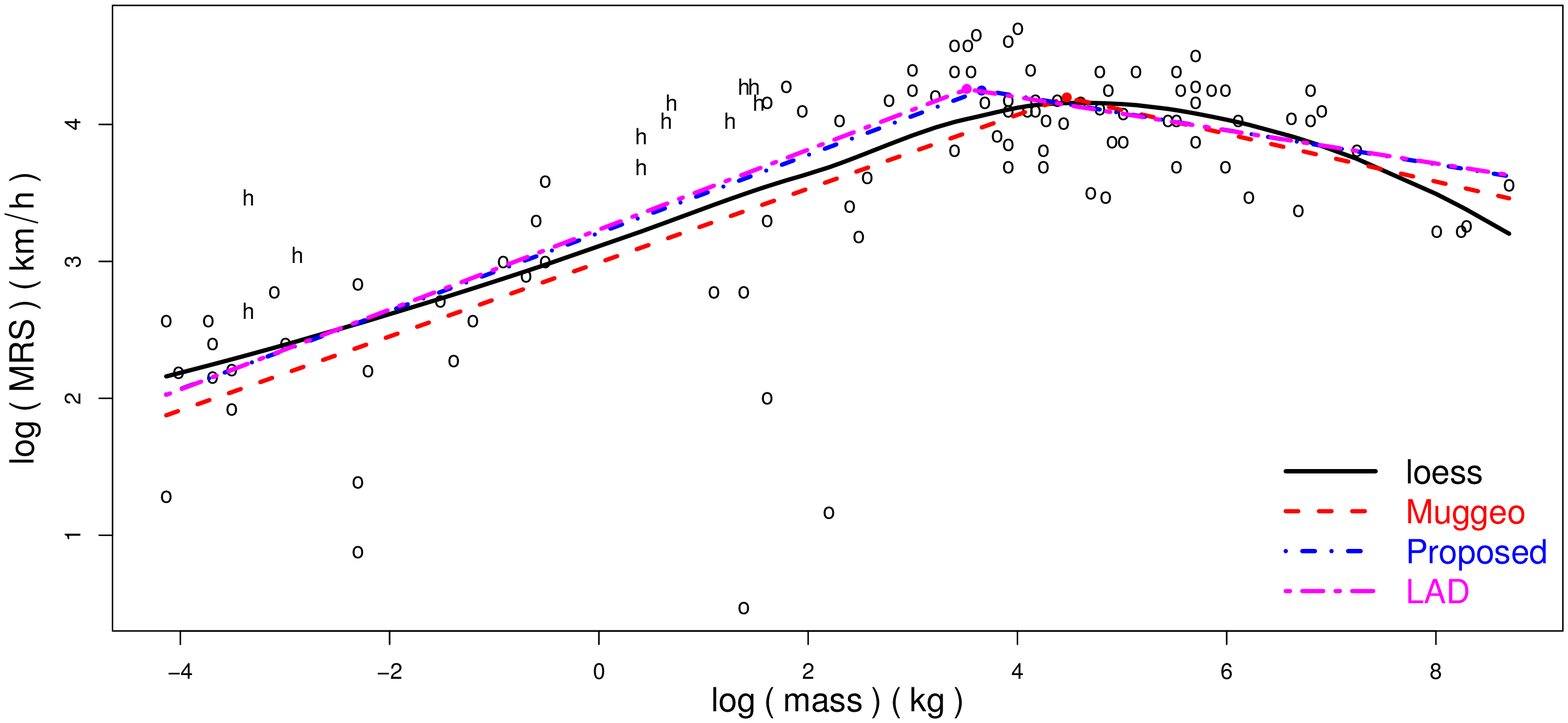}
\label{fig:mrsfit}}
\vspace{5mm}
\subfloat[][Boxplots for the residuals of the three bent line regression methods.]{
\includegraphics[height=3in, width = 0.9\textwidth]{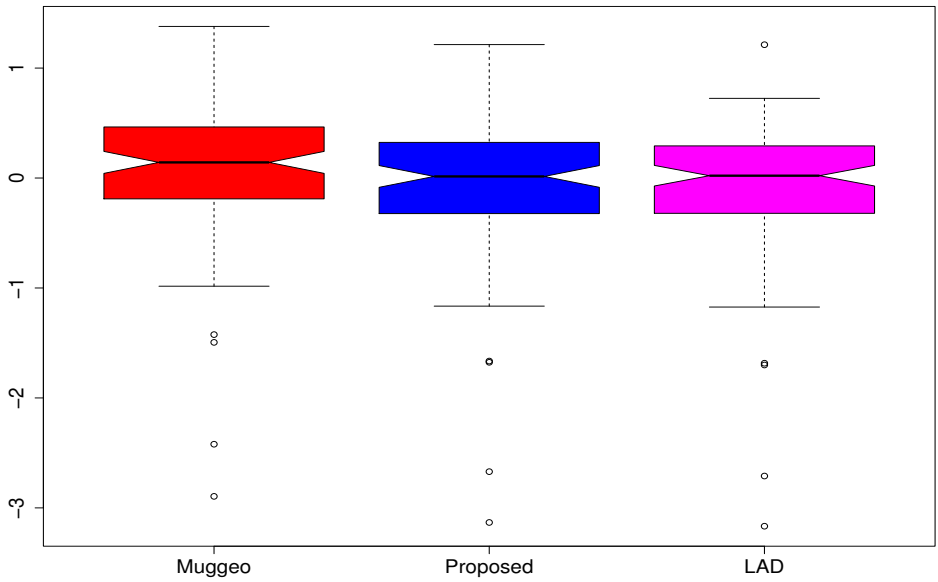}
\label{fig:mrsbox}}
\caption{MRS data analysis. }
\label{fig:mrs}
\end{figure}

Figure~\ref{fig:mrsfit} plots this dataset on the log scale. The animals are labelled according to whether they ambulate by hopping or not, which is believed to affect the running speed. 
The plot indeed shows that there is a slope change in the relation between of log(MRS) and log(mass).    
In addition, it shows that there are several extremely slow animals in the dataset. These animals live in environments where speed is not important for suvival and contribute little to the understanding of how MRS depends on body size. The Grubbs test implies that the three slowest animals ($Y=0.204, 0.470$ and $0.875$) are outliers. This dataset has been analyzed by \citet{li2011bent} using a bent line quantile regression model. To handle these outliers, they focused on the median and higher quantiles.   

Here we analyze this data set using the bent line regression model, 
\begin{eqnarray}\label{mrsmod}
Y_i =\alpha_0
+ \alpha_1 X_i + \beta Z_i +\gamma(Z_i-\tau)_+  + e_i, \quad i=1,...,n, 
\end{eqnarray}
where  $Y_i$ is $\log(MRS)$,  
$Z_i$ is $\log(mass)$, 
$X_i=I(\text{the $i$th mammal is a hopper})$,   
$\tau$ is the change-point location, 
and $e_i$ is the error with an unknown distribution. 
Our test for the existence of a change point shows that the segmented pattern is highly significant (p-value$=0$), which indicates that the estimates and inference from our model are valid. For comparison, we fit the data using our method, Muggeo's method, and bent line quantile regression \citep{li2011bent}.

As shown in Table~\ref{tab:mrs}, all three methods indicate that hopping has a positive effect ($\alpha_1>0$) on MRS. They all report that log(MRS) increases ($\beta>0$) with the increase of log(mass) at first, but then it drops ($\beta + \gamma<0$)  at a certain point. 
However, the estimated change point is somewhat different, at $\exp(3.658) = 38.78$ kg, $\exp(4.472) = 87.53$ kg, and $exp(3.515)=33.6$kg for our method, Muggeo's method and the bent line quantile regression model with $50$th quantile (a.k.a. least absolute deviations regression, LAD), respectively.    
Our estimated coefficients are similar to those of LAD. 
This is unsurprising, as the rank-based regression with the sign scores function $\phi(t)=sgn(t-0.5)$ 
is equivalent to LAD. 
In addition, though all the three methods have similar slopes ($\beta$) before the change point, Muggeo's method has a much lower intercept ($\alpha_0$) than our method and LAD, resulting a lower fitted line. This is likely because Muggeo's method is sensitive to the three outliers with low MRS. 
A close examination of the residuals confirms this conclusion: 
the median of residuals from Muggeo's method has a larger departure from zero than those from our method and LAD (Figure~\ref{fig:mrsbox}).   
This indicates that our method and LAD are much more robust. We performed a five-fold cross validation as in Section \ref{subs:bed} for all the three methods. The prediction error of our method  $(36.959)$ is smaller than those of Muggeo's method $(37.549)$ and the LAD method $(37.243)$.

\begin{table}
	\caption{The estimated parameters and total prediction errors (PE) for MRS data. Their standard errors are listed in parentheses. }\label{tab:mrs}
		\begin{center}
			\begin{tabular}{ccccccc}
				\hline\hline
				& $\alpha_0$ & $\alpha_1$ & $\beta$ & $\gamma$ & $\tau$ & PE \\ 
				Muggeo & 2.991 & 0.841 & 0.270 & -0.444 & 4.472 & 37.549 \\ 
				& (0.078) & (0.189) & (0.024) & (0.092) & (0.445) & \\ 
				Proposed & 3.208 & 0.640 & 0.285 & -0.409 & 3.658 & 36.959 \\ 
				& (0.060) & (0.140) & (0.022) & (0.051) & (0.338) &   \\ 
				LAD & 3.232 & 0.606 & 0.292 & -0.413 & 3.515 & 37.243 \\ 
				& (0.099) & (0.458) & (0.031) & (0.058) &  (0.130) & \\
				\hline
			\end{tabular}
		\end{center}
\end{table}

\section{Discussion}
\label{s:discuss}

In this paper, we developed a rank-based estimation procedure for segmented linear regression model in presence of a change-point. 
By combining a linear reparameterization technique for segmented regression models with rank-based estimation, 
our estimator is both robust against outliers and heavy-tailed errors and is computationally efficient.  
We also proposed a formal testing procedure for the existence of a change point.  
Our results showed that this test is robust while maintaining high power.

There are two interesting extensions of our current work.
First, our work currently is only applicable for detecting one change point. 
It will be interesting to extend it to handle multiple change points.  
When the number of change points is unknown, the estimation and test of the change points would be more complicated.  One possibility is to first determine the number of change points using the idea of the binary segmentation procedures \citet{fryzlewicz2014wild}.  Second,  the linear reparameterization technique \citep{muggeo2003estimating} is applicable to many other segmented linear models with change-points,  
such as generalized linear models and survival models.  It will be worthwhile to extend our robust procedure to these models. 

\section*{Acknowledgments}
This research is partially supported by NIH R01GM109453. 
Zhang's research is partially supported by National Natural Science Foundation of China 
(NSFC) (No.11401194),  
the Fundamental Research Funds for the Central Universities (No.531107050739).

\appendix

\section*{Appendix A}
\label{s:a}

The Appendix contains the technical details of proofs. 

\textit{Regular Conditions}. 
\begin{description}
	\item[(A1)] 
	The density $f$ is absolutely continuous with a bounded first-order derivative and $f>0$.  
	
	\item[(A2)] 
	The design vector satisfies $\mathop{\max}\limits_{1\leq i\leq n}\|W_i\| =o_P(n^{1/2})$ and  
	$\mathop{\lim}\limits_{n\rightarrow \infty} 
	\frac{1}{n} \sum_{i=1}^{n}W_iW_i^T=S_w$ is positive definite matrix. Here, $\|\cdot\|$ is the Euclidean norm. 
	
	\item[(A3)] 
	The change-point $\tau$ lies in a bounded closed interval. 
	
	\item[(A4)]
	The symmetric kernel function $K(\cdot)$ with compact support  $I$ satisfies $\int_I K(u)du = 1$ and has a bounded first derivative. 
	
	\item[(A5)] 
	The bandwidth $h$ satisfies $h\rightarrow 0$ and $nh\rightarrow \infty$ as $n\rightarrow \infty$. 	
\end{description}

We first provide the following convergence results.  
\begin{lemma}\label{lmm1}
	Under the regular conditions,  as $n\rightarrow \infty$, we have 
	\begin{description}
	 \item[(i)] 
	 $S_{wn} \mathop{\longrightarrow}\limits^P S_w$, 
	 
	  \item[(ii)] 
	 $\mathop{\sup}\limits_{\tau}|S_{1n}
	 (\tau) -S_1(\tau)|\mathop{\longrightarrow}\limits^P 0$,  
	 
	  \item[(iii)] 
	  $\mathop{\sup}\limits_{\tau}|\widehat{S}_{1n}
	  (\tau) -S_1(\tau)|\mathop{\longrightarrow}\limits^P 0$,  
		
	  \item[(iv)] 
	  $\mathop{\sup}\limits_{\tau}|S_{2n}(\tau) -S_2(\tau)|\mathop{\longrightarrow}\limits^P 0$. 
	\end{description}
\end{lemma}

\noindent
\textbf{Proof of Lemma ~\ref{lmm1}}. 

For (i), it is easily obtained by using the law of large number. 

For (ii), by the law of large number, $S_{1n}(\tau)\mathop{\longrightarrow}\limits^{P} \mbox{E} S_{1n}(\tau)=S_1(\tau)$ for any given $\tau$. 
Then the uniformly convergence follows with the similar arguments used in Lemma~1 of \cite{hansen1996inference}. 

For (iii), it is sufficient to show that 
$\mathop{\sup}\limits_{\tau}|\widehat{S}_{1n}
(\tau) -S_{1n}(\tau)|=o_P(1)$.   
We can write 
\begin{eqnarray*}
  \widehat{S}_{1n}(\tau)- S_1(\tau)
 &=&
 \frac{1}{n}
 \sum_{i=1}^{n}\sqrt{12}\left[
 \widehat{f}(\widehat{e}_i)-f(\widehat{e}_i)
 \right]
  \bW_i (Z_i-\tau)I(Z_i\leq \tau)  \\
 &&+
 \frac{1}{n}
 \sum_{i=1}^{n}\sqrt{12}\left[
 f(\widehat{e}_i)-f(e_i)
 \right]
 \bW_i (Z_i-\tau)I(Z_i\leq \tau)  \\
 &&+
 S_{1n}(\tau)-S_1(\tau) \\
  &\equiv&
   I_1+I_2+I_3.
\end{eqnarray*}
Clearly, $\sup\limits_{\tau}|I_1|=o_P(1)$ by the uniform convergence of the kernel density estimator. 

Note that 
\begin{eqnarray*}
	|I_2| 
	&\leq&   	
	\frac{1}{n}
	\sum_{i=1}^{n}\sqrt{12}
	\left|\bW_i (Z_i-\tau)I(Z_i\leq \tau) \right|
	\max_i\left|
    f(Y_i-\widehat{\bxi}^T\bW_i)
	-f(Y_i-\bxi^T\bW_i)
	\right|. 	
\end{eqnarray*}
By the Conditions (A4) and (A5), and $\|\widehat{\bxi}-\bxi\|=O_P(n^{-1/2})$ in the proof of Theorem~\ref{thm1}, and the mean-value theorem,  we get 
\begin{eqnarray*}
	\max_i\left|
	f(Y_i-\widehat{\bxi}^T\bW_i)
	-f(Y_i-\bxi^T\bW_i)
	\right|
	\leq \max_i \|\bW_i\|\cdot |f'(\zeta^T\bW_i)|\cdot	 \|\widehat{\bxi}-\bxi\|
	=o_P(1),
\end{eqnarray*}
where $\zeta$ lies in the segment between $\widehat{\bxi}$ and $\bxi$.  Thus, $\sup\limits_{\tau}|I_2|
= o_P(1)$. 

Furthermore,  $\sup\limits_\tau|I_3|=o_P(1)$ follows from (ii), and hence (iii) holds.  

The proof of (iv) is similar to that of (ii) and is omitted here.  
\qed

\medskip
\noindent
\textbf{Proof of Theorem~2.1}

Note that 
\begin{eqnarray*}
\widehat{\bxi}
\equiv
(\widehat{\balpha}, \widehat{\beta}) 
=
\arg\min_{\balpha, \beta}
\sum_{i=1}^{n} \sqrt{12}\left(\frac{R(Y_i-\bxi^T\bW_i)}{n+1}-0.5\right) \times\left(Y_i-\bxi^T\bW_i\right),
\end{eqnarray*} 
which is equivalent to solve the estimating equation, 
\begin{eqnarray*}	
	S_n(\bxi) 
	&=&
	-\frac{d}{d\bxi}
	\sum_{i=1}^{n} \sqrt{12}\left(\frac{R(Y_i-\bxi^T\bW_i)}{n+1}-0.5\right)\times\left(Y_i-\bxi^T\bW_i\right)\\
	&=& 
	\sum_{i=1}^{n} \sqrt{12}\left(\frac{R(Y_i-\bxi^T\bW_i)}{n+1}-0.5\right)\bW_i.
\end{eqnarray*} 
Under the local alternative model~ (8) , 
that is, 
\begin{eqnarray*}
Y_i = \balpha^T \bX_i  + \beta Z_i + n^{-1/2}\gamma(Z_i-\tau)_+ + e_i,\quad i=1,...,n, 
\end{eqnarray*} 
we have 
\begin{eqnarray*}
	S_n(\bxi)
	&=& 
	\sum_{i=1}^{n} \sqrt{12}\left[
	\frac{R\{e_i+n^{-1/2}\gamma(Z_i-\tau)_+\}}{n+1}-0.5
	\right]\bW_i\\
	&=&
	\sum_{i=1}^{n} \sqrt{12}\left[
	\frac{n}{n+1}F_n\left(e_i+n^{-1/2}\gamma(Z_i-\tau)_+
	\right)-0.5\right]\bW_i\\
	&=& 
	\sum_{i=1}^{n} \sqrt{12}\left[
	F\left(e_i+n^{-1/2}\gamma(Z_i-\tau)_+\right)-0.5
	\right]
	\bW_i+o_P(1)\\
	&=& 
	\sum_{i=1}^{n} \sqrt{12}\left[
	F(e_i) -0.5 + f(e_i) n^{-1/2}\gamma(Z_i-\tau)_+
    \right]\bW_i, 
\end{eqnarray*}
where the last equality is followed by Taylor expansion. 

By the Theorem~A.3.8 in \cite{hettmansperger2011robust}, 
it yields that 
\[
	n^{-1/2}S_n(\widehat{\bxi})=n^{-1/2}S_n(\bxi)-
	\frac{1}{c_\phi}\left(\frac{1}{n}\sum_{i=1}^{n}\bW_i \bW_i^T\right) \sqrt{n}(\widehat{\bxi}-\bxi)+o_P(1). 
\]
Note that $	n^{-1/2}S_n(\widehat{\bxi})=0$, and by Lemma~\ref{lmm1}, it follows that 
\begin{eqnarray*}
	\sqrt{n}\left(\widehat{\bxi}-\bxi\right)
	&=&
	c_\phi S_w^{-1}n^{-1/2}\sum_{i=1}^{n} \sqrt{12}
	\left[F(e_i)-0.5\right]W_i \\
	&&
	+ c_\phi S_w^{-1}n^{-1/2}\sum_{i=1}^{n} \sqrt{12}
	\gamma n^{-1/2} f(e_i)(Z_i-\tau)_+ W_i +o_P(1).  
\end{eqnarray*}
Now, under the local alternative model  (8),  
we can write $R_n(\tau, \widehat{\xi})$ as 
\begin{eqnarray*}
 && 
 R_n(\tau, \widehat{\bxi})\\
 &=&
 \frac{1}{\sqrt{n}}\sum_{i=1}^{n} \sqrt{12}
 \left[
 \frac{R\left(e_i-(\widehat{\bxi}-\bxi)^T
 	\bW_i + n^{-1/2}\gamma (Z_i-\tau)_+\right)}{n+1}-0.5
 \right]
 \left(Z_i-\tau\right) I(Z_i\leq \tau)\\
 &=&
  \frac{1}{\sqrt{n}}\sum_{i=1}^{n} \sqrt{12}
  \left[
  \frac{n}{n+1}
  F_n\left\{
  e_i-(\widehat{\bxi}-\bxi)^T
  \bW_i + n^{-1/2}\gamma (Z_i-\tau)_+
  \right\}-0.5
  \right]
  \left(Z_i-\tau\right) I(Z_i\leq \tau)\\
 &=& 
   \frac{1}{\sqrt{n}}\sum_{i=1}^{n} \sqrt{12}\left[F(e_i)-0.5
   -f(e_i)(\widehat{\bxi}-\bxi)^T \bW_i 
   + n^{-1/2} f(e_i)\gamma (Z_i-\tau)_+\right] 
   \left(Z_i-\tau\right) I(Z_i\leq \tau), 
\end{eqnarray*}
where the last equality is used Taylor expansion. 

By plugging in the representation for $\sqrt{n}(\widehat{\bxi}-\bxi)$ and some algebraic manipulation, we have 
\begin{eqnarray*}
	R_n(\tau, \widehat{\bxi}) = 
	n^{-1/2}\sum_{i=1}^n \sqrt{12} \bigg[F(e_i)-0.5\bigg]
	\bigg[
	(Z_i-\tau)I(Z_i\leq \tau)- 
	c_\phi S_1(\tau)S_w^{-1}W_i
	\bigg] 
	+ q(t) +o_P(1). 
\end{eqnarray*}  
The  remainder conclusion for weak convergence of $R_n(t,\widehat{\bxi})$ is easily obtained by following the proofs in \cite{stute1997nonparametric}. 
\qed

\medskip
\noindent
\textbf{Proof of Theorem~2.2}

The proof follows the same line as that for Theorem ~2.1,  
then it is omitted for saving space.     
\qed

\medskip
\noindent 
\textbf{Proof of Theorem~2.3}

We divide the proof into three steps. 

First, we show that the covariance function of $R_n^*$ converges to that of $R$. Define 
\[
	R_n^{**}(\tau)=n^{-1/2}\sum_{i=1}^{n} u_i\sqrt{12}\bigg[F_n(e_i)-0.5\bigg]	
	\bigg[
	(Z_i-\tau)I(Z_i\leq \tau)- 
	c_\phi S_1(\tau)S_{w}^{-1}W_i
	\bigg].
\]
By the fact that the uniformly convergence of $\widehat{F}_n(\cdot)-F_n(\cdot)$ and $\widehat{c}_\phi- c_\phi$,  along with the uniform convergence of $\widehat{S}_{1n}(\tau)-S_1(\tau)$ in Lemma~\ref{lmm1}, 
we can easily show $R_n^*(\tau)$ and $R_n^{**}(\tau)$ are asymptotically equivalent in the sense that 
\[
	\sup_\tau \|R_n^*(\tau)-R_n^{**}(\tau)\|=o_P(1).
\]
Note that $u_i$'s are independent of $(Y_i,\bX_i, Z_i)$,  
and $\mbox{E}u_i=0$, $\mbox{Var}(u_i)=1$.   
Then, for any $\tau_1, \tau_2$, 
the covariance function of $R_n^{**}$ is 
\begin{eqnarray*}
  &&	
  Cov\left(R_n^{**}(\tau_1), R_n^{**}(\tau_2)\right)\\
  &=&
  \frac{1}{n}\sum_{i=1}^{n}
 \mbox{E}\bigg(u_i^2 
  12 \left[F(e_i)-0.5\right]^2
  \left\{
  (Z_i-\tau_1)I(Z\leq \tau_1) -c_\phi S_1(\tau_1)^TS_w^{-1}W
  \right\}\\
  && \times
  \left\{
  (Z_i-\tau_2)I(Z_i\leq \tau_2) -c_\phi S_1(\tau_2)^TS_w^{-1}W
  \right\}\bigg)\\
  &=&  
  \mbox{E}
  	\bigg[
  	\left\{
  	(Z-\tau_1)I(Z\leq \tau_1) -c_\phi S_1(\tau_1)^TS_w^{-1}W
  	\right\}\cdot
  	\left\{
  	(Z-\tau_2)I(Z\leq \tau_2) -c_\phi S_1(\tau_2)^TS_w^{-1}W
  	\right\}
  	\bigg]. 
\end{eqnarray*}
which  is the same as the covariance of $R$. 

Second, it is easily to show that any finite-dimensional projection of $R_n^*(\tau)$ converges to that of $R(\tau)$, 
by the central limit theorem.

Third, $R_n^*(\tau)$ is uniformly tight.  
Note that the class of all indicator functions 
$I(Z\leq \tau)$ is a  Vapnik-Chervonenskis (VC) class of functions. 
Then, the class of functions 
$$
\mathcal{F}_n=\left\{
(Z_i-\tau)I(Z_i\leq \tau)- 
c_\phi S_{1n}(\tau)S_w^{-1}W_i: 
\tau \in R^1 
\right\}
$$ 
is a VC class of functions. Thus, by the equicontinuity lemma~15 of \citep{pollard1984convergence}, 
one can show that $R_n^*(\tau)$ is uniformly tight. 
Then, by the Cramer-Wold device, the proof of Theorem~2.3 
is completed. 
\qed

\section*{Appendix B}
\label{s:b}

This Appendix provides the algorithm for  testing the existence of a change-point via the wild bootstrap method based on Muggeo's method.

Similarly, 
the test statistic based on the Muggeo's  segmented regression is given by 
\[
\widetilde{T}_n =\sup_{\tau\in T} \left|\widetilde{R}_n (\tau, \widetilde{\bxi}) \right|
\]
where 
\[
\widetilde{R}_n(\tau, \widetilde{\bxi})= \frac{1}{\sqrt{n}}\sum_{i=1}^{n} 
\left(
Y_i-\widetilde{\bxi}^T\bW_i\right) 
\left(Z_i-\tau\right) I(Z_i\leq \tau),
\]
where $\widetilde{\bxi}$ is obtained by  Muggeo's method under the null hypothesis. 

The algorithm for the wild bootstrap method based on Muggeo's  method is summarized as follows.

\fbox{
\begin{minipage}{\dimexpr\textwidth-2\fboxsep-2\fboxrule\relax} 
	\parbox{1\textwidth}{%
		\textbf{Algorithm~3}:
		\begin{center} 
\begin{description}
	\item[Step~1]  
	Generate iid $\{u_1,\cdots, u_n\}$  with $u_i=v_iw_i$, where $v_i$ is generated from the standard normal distribution  $N(0,1)$, and $w_i$ (independent of $v_i$'s) from the two-point mass distribution with equal probability at $1$ and $-1$.  
	
	\item[Step~2]  
	Calculate the test statistic 
	\[
	\widetilde{R}_n^*(\tau)= \frac{1}{\sqrt{n}}\sum_{i=1}^{n} u_i	
	\bigg[
	(Z_i-\tau)I(Z_i\leq \tau)- 
	 \widetilde{S}_{1n}(\tau)S_{wn}^{-1}W_i
	\bigg],
	\]
	where 
	$$
	\widetilde{S}_{1n}(\tau)= \frac{1}{n}\mathop{\sum}\limits_{i=1}^{n} W_i(Z_i-\tau)I(Z_i\leq \tau). 
	$$

	\item[Step~3] 
	Repeat Steps~1--2 with $NB$ times to obtain $\widetilde{T}_{n}^{*(1)},\cdots, \widetilde{T}_n^{*(NB)}$. 
	Calculate the p-value as  
	$\widetilde{p}_n =\frac{1}{NB}\mathop{\sum}\limits_{j=1}^{NB} I\{\widetilde{T}_n^{*(j)}\geq T_n\}$. 	
\end{description}
	\end{center}
	}%
\end{minipage}
}%

\section*{References}
\bibliographystyle{elsarticle-harv} 
\bibliography{robustSeg_arxiv_20160607}





\end{document}